\documentclass{article}
\usepackage{epsfig}
\usepackage{dcolumn}
\usepackage{bm}
\usepackage{amssymb}
\usepackage{amsmath}
\usepackage{amsfonts}
\usepackage{graphicx}
\usepackage{psfrag}

\title{A Computational Anthropic Principle: Where is the Hardest Problem in the Multiverse?}
\author{Navin Sivanandam}

\begin{document}
\begin{flushright} 
UTTG-14-09\
\end{flushright}
\begin{center}
\Large{\bf A Computational Anthropic Principle: Where is the Hardest Problem in the Multiverse?}\\
\vspace{10pt}
\large{\bf Navin Sivanandam}\\
\small{email: navin.sivanandam@gmail.com}\\
\vspace{10pt}
\small{\it Theory Group, Department of Physics, University of Texas, Austin, TX 78712}
\end{center}
\begin{center}
{\bf Abstract}\\
\end{center}
\begin{quote}
The anthropic principle is an inevitable constraint on the space of possible theories. As such it is central to determining the limits of physics. In particular, we contend that what is ultimately possible in physics is determined by restrictions on the computational capacity of the universe, and that observers are more likely to be found where more complicated calculations are possible. 

Our discussion covers the inevitability of theoretical bias and how anthropics and computation can be an aid to imposing these biases on the theory landscape in a systematic way. Further, we argue for (as far as possible) top-down rather than bottom-up anthropic measures, contending that that the latter can often be misleading.

We begin the construction of an explicit computational measure by examining the effect of the cosmological constant on computational bounds in a given universe, drawing from previous work on using entropy production as a proxy for observers by Bousso, Harnik, Kribs and Perez. In addition, we highlight a few of the additional computational considerations that may be used to extend such a measure.
\end{quote}
\subsection*{Die Grundlagen der Physik}
As he was wont to do rather frequently, in 1928 David Hilbert outlined a list \cite{Hilbert} and a program\footnote{Actually pinning a single date on Hilbert's program is not entirely accurate, since the work in question developed over several decades. On the other hand, it makes for better story-telling.}. The program in question was the formalization of mathematics and mathematical reasoning; Hilbert argued that the primary aim of researchers should be to construct an axiomatic foundation of mathematics with the following three desiderata:
\begin{description}
\item[Consistency] The system of axioms should be consistent and provably so.
\item[Completeness] All mathematical truths (at least in principle) be deduced from those axioms.
\item[Decidability] There should be a well-defined procedure such that given any statement in mathematics, the truth of that statement can be established within a finite time.
\end{description}
As Hilbert himself expressed some decades earlier\footnote{Problem number 6 on the famous 1900 set of 23.} it would be equally desirable to make some progress axiomatizing physics. In some ways this is both an easier and harder problem: easier because one has the real world to guide one's choice of axioms, harder because, unlike mathematics, it's not at all clear what physics is. While axiomatizing physics is tricky and barely defined, the lack of a question is rarely a sufficient blockade to physicists searching for answers.

So, what should we require from a set of physical axioms? Well, at the very least, since we conventionally describe the physical world in the language of mathematics, we should insist on Hilbert's criteria as a minimum. Unfortunately, we famously run into trouble here since thanks to G\"odel \cite{Godel}, Church \cite{Church} and Turing \cite{Turing}, we know that whatever axioms we choose to use in mathematics, we can expect neither consistency, completeness nor decidability. Having said that, however, there is a real sense in which the above desiderata are still, well, desirable for physics. We want our axioms (and thus our physical theories) not to be mutually inconsistent; we would like, in principle, to be able to make as many predictions (engender as many explanations) about the world as we can; and we would further hope that predictions are also easy in practice, i.e. we should be able to compute as many consequences of our theory as possible. It is this last issue, computation, that we will focus on here.

We have given the Entscheidungsproblem (the decision problem) a slightly different meaning when we apply it to physics. Rather than asking if the veracity of every statement can be established, instead we ask what the most complex question that can be asked and answered is in a given physical theory. Moreover, we consider theories to be more desirable they permit calculations of greater complexity\footnote{Which is at least consistent with our evaluation of the desirability of other commodities in life.}. But why?

The why is perhaps best motivated by asserting that good physical theories have a further desideratum:
\begin{center}
``\emph{Theories of physics should maximize the number of physicists\footnote{It is conventional to use the term observer here, we will use both interchangeably in order emphasize that more than just observation is needed.}.}''
\end{center}
This is nothing more than a restatement of the increasingly in vogue (Weak) Anthropic Principle: The universe should be consistent with our existence. The Anthropic Principle has a storied history \cite{Barrow} and at least as many detractors as adherents. We will delve a little into this murky territory below, but in general we take the point of view that finding satisfying theories of the world usually requires us to define a measure on the space of theories (more prosaically we must decide what makes a theory satisfying), and that anthropic reasoning is simply a route to such a metric. Moreover, rather than thinking in terms of people, observers or other such involved entities, we assert that anthropic measures are best expressed in terms of computation.

A related point of view has been espoused by Bousso et al. \cite{Bousso:2007kq}, who consider anthropic constraints by studying the production of entropy during the universe's history. Our goal here is to expand upon this idea by explicitly advocating computation as anthropics, and though we will end up mostly considering entropy, we believe that a computational point of view can be considerably more constraining.

Let us begin with a description of our extended cosmological playground. Before we do that, though, a few words of warning.

\subsection*{Caveat Lector}
The universe is not a computer. Or rather, while viewing physics as digital process (e.g. \cite{Wolfram} and \cite{Lloyd}) is not unheard of (and perhaps not even unreasonable), that is not the point of view taken here\footnote{Caveating the caveat, however, Feynman's musings \cite{Feynman} on this subject are hard to find fault with, and perhaps offer some direction on how one might use notions of computation to constrain less ``human'' aspects of physical law: ``It always bothers me that, according to the laws as we understand them today, it takes a computing machine an infinite number of logical operations to figure out what goes on in no matter how tiny a region of space, and no matter how tiny a region of time. How can all that be going on in that tiny space? Why should it take an infinite amount of logic to figure out what one tiny piece of space/time is going to do?''}. Rather, it is the processes of physicists that are essentially computational in nature. Moreover, the existence and the limitations (and perhaps even the predilections) of physicists are (at least for the purposes of this essay) an intrinsic part of subject\footnote{Though, let us stress that talk of hermeneutics \cite{Sokal} would probably be taking things too far.}.

\subsection*{A Landscape of Postmodern Physicists}
Much of fundamental physics can be viewed as a quest for a Theory of Everything. This is a far less lofty goal than it sounds, for such a theory is not expected to describe everything, but rather should contain within it a framework for deducing appropriate effective theories for any given phenomenon. If something like this is true, then we can be confident that our various descriptions of different parts of the world are mutually consistent. Much to the delight of budding physicists and pop science publishers, finding such a theory is very much work in progress. However, there are candidates, among which is string theory.

String theory has many desirable features, many amazing properties and is almost infuriatingly rich. While the details of the theory don't concern us, it's richness does. The theory describes the physics of very high energies, and has the remarkable property that when one tries to inquire about what we should see in the relatively low energy world where we live (and where one might expect to find cosmologies), one finds a vast number of possibilities, each with different physics. This rich set of possibilities is known as the ``Landscape'' \cite{Susskind:2003kw}. While the string landscape provides much of the motivation for discussing a space of theories (and provides the model we use in this essay), it is worth pointing out that this notion arises even in more well established constructions such as the Standard Model; either simply through the familiar parameter space or through more involved scenarios \cite{ArkaniHamed:2007gg}.

The precise nature of landscape is debatable \cite{Dine:2004fw}; it's unclear how many vacua (low energy worlds) there are, how these vacua are populated and how the laws of physics vary between vacua. This uncertainty makes finding a measure a thorny task, so we will have to make some assumptions in order to progress. In particular:
\begin{itemize}
\item Gravity survives in low energy vacua.
\item The only relevant vacua are those which give rise to FRW (Friedmann-Robertson-Walker) universes -- these have homogenous, isotropic cosmologies.
\item There is a spectrum of possible values for the cosmological constant (cc or $\Lambda$) in our possible universes. The cc corresponds to the intrinsic energy density of empty space; in our universe it is approximately $10^{-120}M_p^4$ where $M_p\sim10^{19}$ GeV is the Planck mass (we will mostly use units such that $M_p=1$ along with $k_B=c=\hbar=1$). Further, we assume that this $\Lambda$ is positive and lies in a range $\Lambda_{\text{min}}\leq\Lambda\leq\Lambda_{\text{max}}$.
\item Inflation takes place and produces the initial spectrum of density perturbations. This also implies that the universe is spatially flat and at the critical density.
\end{itemize}
From an anthropic point of view, these aren't necessarily good assumptions. However, as galling as it may be, we have no choice if we wish to progress. With this vast (if slightly constrained) landscape at our disposal, we have a number of avenues via which we may explore the limits of physics. Cosmology is perhaps the easiest of these avenues to negotiate, since it allows us to say something about the fundamentals of computation in a particular vacuum without detailed knowledge of the physics.

With a concrete model of the space of theories the anthropic principle may seem redundant, after all why should we introduce additional limits on what is possible? Well, despite its dreadfully postmodern tone, the claim that the existence of physicists is a necessary condition for a physical universe is undeniably true\footnote{Stipulating that the unobservable is unphysical.}. Moreover, one cannot pick from a space of theories without applying a measure, something even the most staunchly unanthropic of scientists do. Consider, if you will, the problem of fine-tuning.

Nature is described by models, these models contain both dynamics and parameters. For example, in the Standard Model we must input various numbers (determined by experiment) in order to have a complete description of particle interactions. Measuring these numbers is not sufficient, we also seek to explain why they have the values they have. In particular, we are troubled by quantum corrections to parameters that are much larger than their measured values. Such situations imply a precise cancelation between the bare value of a parameter and its quantum corrections, i.e. we have \emph{fine-tuning}. While this seems intuitively problematic, one should note that ones discomfort is entirely a theoretical prejudice against fine-tuning -- we hold an a priori belief that such a precise cancelation is unlikely. Amongst the canonical examples of fine-tuning, the worst is the cosmological constant, where we calculate quantum corrections of order $M_p^4$, but only measure $\Lambda\sim10^{-120}M_p^4$.

The anthropic principle provides an additional (and well-justified) theoretical bias to help us navigate our expectations of theory space. This is not say that all is well with blithely diving into anthropic arguments; they may be obviously true, but using them to extract meaningful information is not straightforward. It is far from clear what is the relevant definition of, for example, ``typical observer''. We concern ourselves with the observer part here, but there are deep issues (and much disagreement) with interpretations of typicality \cite{Bostrom:2007zz}-\cite{Page:2007bt}.

To apply our computational interpretation of the anthropic principle, consider the following probability measure on the space of observed theories:
\begin{align*}
P\left(\text{Theory is observed}\right) &= P\left(\text{Theory}|\text{Observers}\right)\\
&= \frac{P\left(\text{Observers}|\text{Theory}\right)P\left(\text{Theory}\right)}{P\left(\text{Observers}\right)}\\
&= N\ P\left(\text{Computational Capacity}|\lambda_i\right)P\left(\lambda_i\right)
\end{align*}
Here the first line expresses the anthropic principle, the second is Bayes' formula and the third parameterizes theory space by some set of variables $\lambda_i$, adds a normalization factor $N$ and asserts that the probability distribution of observers goes like the computational capabilities of a given theory. $P\left(\lambda_i\right)$ is the (prior) probability distribution of the parameters $\lambda_i$, given to us by string theory, or some other candidate Theory of Everything. In what follows our goal is to compute an estimate for the quantity $P\left(\text{Computational Capacity}|\lambda_i\right)$.

Actually, we've cheated a little here and missed out what is possibly the most significant part of our probability distribution -- the weighting from eternal inflation (see e.g. \cite{Guth:2007ng} for summary). This is the process through which the landscape is populated by the nucleation of new vacua from existing ones \cite{Coleman:1980aw}, \cite{Brown:1987dd}. Eternal inflation is a hard problem, in which there has been much debate, but little consensus. We also note that finding the prior distribution $P\left(\lambda_i\right)$ is far from straightforward, though has been some success \cite{Bousso:2000xa}, \cite{Denef:2004ze}, particularly when focussing on the cosmological constant alone. A variety of arguments suggest that the prior distribution for $\Lambda$ is uniform over ranges that do not include $\Lambda=0$ \cite{Weinberg:2000qm}, but our forays in the landscape are far from extensive and there may be more to say. As such our comments about values for $\Lambda$ implied by anthropics alone should be understood as valid only in the limited scenarios where neither eternal inflation nor the prior distribution have much effect.

\begin{figure}[ht]
\center \includegraphics[height=60mm]{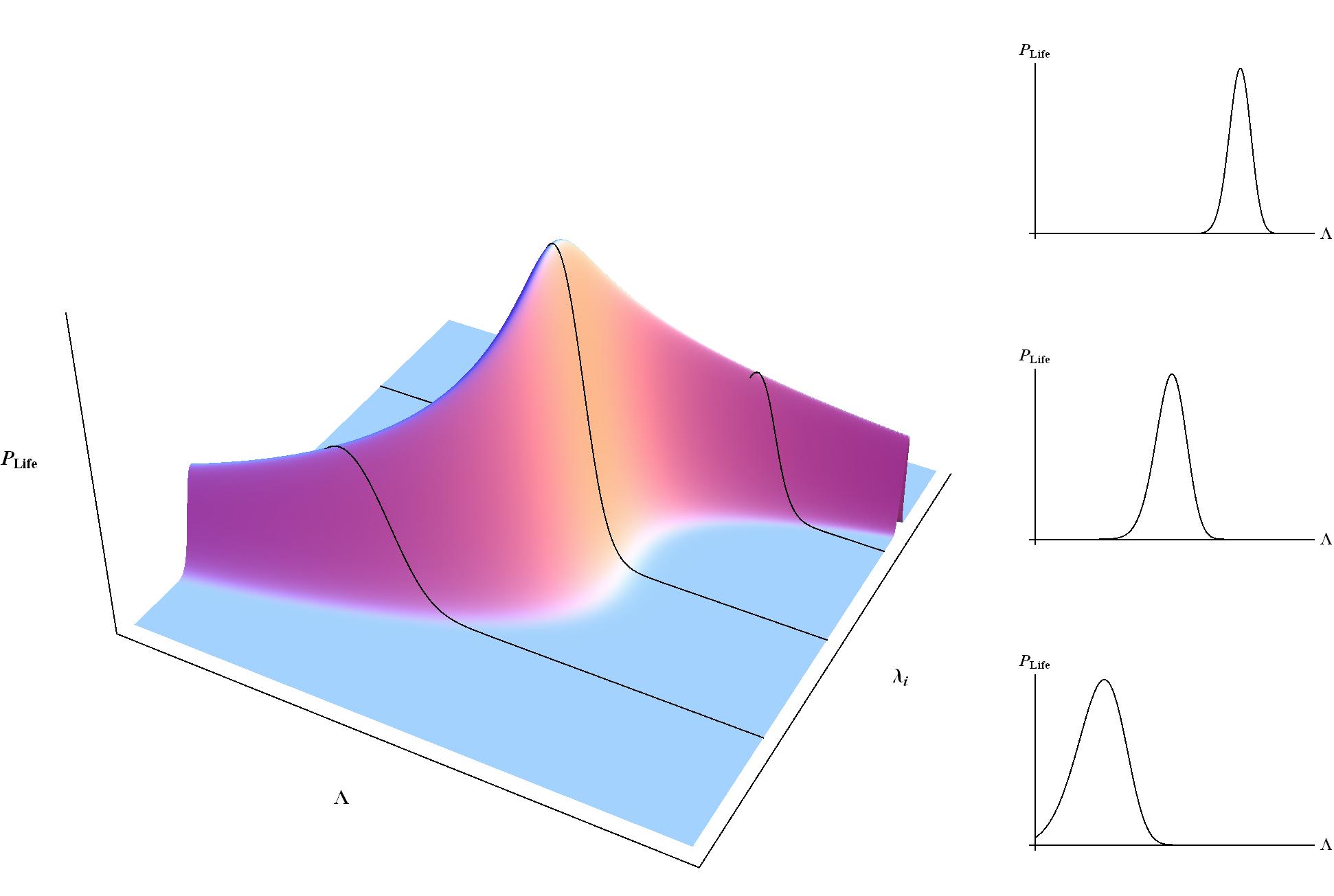}
\caption{\textit{A complicated potential can lead to vastly different probability distributions for $\Lambda$ depending on what values we fix for the $\tilde{\lambda_i}$.}}
\label{prob}
\end{figure}

Anthropic limits for the cosmological constant are manifold and a wide variety of proxies have been used for observers -- from the existence of collapsed structure \cite{Weinberg:1987dv} to the presence of large amounts of entropy \cite{Bousso:2007kq}. However, these  bounds are constructed from the bottom up -- i.e. one starts by fixing many of the parameters of our universes (often all but the one whose value we are trying to anthropically explain) and then conditioning on some proxy for observers. Such an approach has issues. To begin with, it suffers from the propensity to favor observers that are too much like us (the so-called ``carbon bias''). Along with this, many facets of the underlying description of the universe conspire together to give the phenomena that we need for life (such as structure), and holding them constant does not, in general, allow one to find the maximum of the probability distribution -- Figure \ref{prob} provides a visual representation of this. More concretely, a bottom up approach involves claiming the probability distribution of observers is given by:
\begin{equation*}
P\left(\text{Observers}|\Lambda\right)=\left.P\left(\text{Observers}|\lambda_i\right)\right|_{\tilde{\lambda_i}=\text{constant}}
\end{equation*}
Where $\tilde{\lambda}_i$ represents all the parameters of theory excluding $\Lambda$. This is a mostly unjustified assertion (see \cite{ArkaniHamed:2005yv} for some justification), implying either that the functional form for the $\Lambda$ dependence of $P\left(\text{Observers}\right)$ is independent of other parameters or that every $\tilde{\lambda_i}$ has a value strongly picked out by some non-anthropic mechanism.

An obvious counter-argument to the case against bottom-up anthropics would be to point out that this is how we do physics in laboratory experiments; we hold everything fixed and try to understand how changing one variable affects a system. However, anthropic reasoning is very different from experimental reasoning. We are not trying to explore the possibilities of our universe, but rather those of the multiverse -- \emph{we do not know that the majority of observers are close to us in parameter space, and we should not reason as if that is the case}.

The distinction between top-down and bottom-up anthropics highlights what is perhaps the greatest advantage of thinking about the anthropic principle in terms of computation. It is both easier to avoid carbon bias (the fundamental qualities of observers and observations lend themselves to expression in the language of computing) and it is considerably easier to construct top-down anthropic measures, since many of the computational properties of a theory can be understood without a detailed derivation of its phenomenological consequences.

\subsection*{When in Doubt, Calculate!}
As much as we prefer elegant formulae and pithy expressions, ultimately physics comes down to the numbers we measure and the numbers we predict. Prediction usually involves taking some subset of the data we already have and using it alongside the theory to calculate some number corresponding to a future measurement. Our ability to calculate depends on two things: the computational complexity of the calculation and the power of the computer (animal, mineral or otherwise).

Computational complexity is a rich subject with deep unsolved problems\footnote{Consider this the customary notice that there's a million dollars in it if you figure them out \cite{Clay}.} and many implications for physics. One might use notions of complexity to bound what is and isn't possible in the physical world \cite{Aaronson:2005qu} or to limit landscape explorations \cite{Denef:2006ad}.

A physical theory is not only fettered by the difficulty of the calculations it foists upon unsuspecting graduate students, but also by the propensity of said graduate students to exist in the world(s) predicted by said theory. Sadly, deducing the existence and capabilities of graduate students from a Theory of Everything is a little tricky\footnote{Though perhaps there are ambitious graduate students out there who would like to try?}, so instead we use a proxy: how easy is it to perform a computation?

As is now well appreciated, there are deep connections between physics and information. We do not have space even to cover the highlights (the reader is referred to chapter one of Preskill's quite excellent lecture notes \cite{Preskill}), but there is one particular insight that we should note:
\begin{description}
\item[Landauer's Principle] Due to Rolf Landauer (1961), this states that the erasure of information is dissipative and is associated with a change of entropy, $\Delta S=k_B\ \text{ln}\ 2$.
\end{description}
$k_B$ is the Boltzmann constant; we will work with units such that this is equal to 1 (the factor of $\text{ln}\ 2$, along with others of a similar size, will generally be ignored). This principle is particularly relevant for cosmological bounds on computing, where, as we will see, we do not have an unlimited amount of storage space. For the moment we will concentrate on constraints imposed on the numbers of bits and on the number of operations (ops). These are both necessary for computation, but clearly not sufficient, and before we draw this essay to a close we will speculate on other possible computational constraints.

As with all interesting ideas, the physical limits on computation in our universe have already been studied by various authors. Lloyd \cite{Lloyd:2001bh} calculates the computational capacity of our universe (though using matter-dominated cosmology, rather than the cosmological constant-dominated universe -- the $\Lambda$CDM model -- that we now believe we live in), by placing constraints on both the number of bits and the number of ops. A similar, but slightly more refined calculation, is carried out by Krauss and Starkman \cite{Krauss:2004jy} for the $\Lambda$CDM cosmology. Much of what follows using similar techniques, albeit in a different spirit.

Constraining bits follows from noting that the information storage capacity, $I$, of a system is related to the maximum entropy ($S$):
\begin{equation*}
I=\frac{S}{k_B\ \text{ln}\ 2}\sim S
\end{equation*}
Hence, the maximum information processing capability is given by the maximum entropy of the system. In our cosmological constant-dominated universe we use the holographic principle \cite{'tHooft:1993gx}, \cite{Susskind:1994vu} to bound the maximum entropy of a gravitational system by the area of a suitable bounding surface (detailed definitions vary, but the differences are unimportant for us). Every FRW cosmology with a positive cosmological constant evolves in the far future to de Sitter space -- an exponentially-expanding spacetime with a cosmological horizon. Thus the amount of space we can observe is limited, no matter how long we wait. The distance to the horizon is set by the rate of expansion of the universe, the Hubble constant, which is in turn a function of $\Lambda$ ($M_p=1$, $O(1)$ numbers ignored):
\begin{equation*}
\text{distance to horizon}\sim\frac{1}{\Lambda^{1/2}},\Rightarrow \text{area of horizon}\sim\frac{1}{\Lambda}
\end{equation*}
This bounds the information processing ability as:
\begin{equation*}
I\lesssim\frac{1}{\Lambda}
\end{equation*}
Of course, we may be (and in some sense must be) far from saturating this bound. However, it is striking that for a wide class of cosmologies we can limit this aspect of the computational power with reference only to the value of $\Lambda$, a parameter that we should feel confident we can extract from a fundamental theory. As previously mentioned, entropy has been used as proxy for observers in the work of Bousso et al. \cite{Bousso:2007kq}, and while the focus is on bottom up anthropic constraints the authors have also noted the possible existence of a top down bound \cite{Bousso:2006ucsb}.

How about the number of ops? Here, both Lloyd and Krauss and Starkman use the Margolus/Levitin theorem \cite{Margolus:1997ih}, which states the number of ops per second is limited by the energy of the system:
\begin{equation*}
\text{ops/sec}\leq\frac{2E}{\pi\hbar}\Rightarrow \text{ops/Planck time}\sim E
\end{equation*}
To find a bound in de Sitter space is straightforward -- the value of $\Lambda$ sets both the energy density ($=\Lambda$) and the volume ($=1/\Lambda^{3/2}$), giving ($t_p$ is the Planck time):
\begin{equation*}
\text{ops}/t_p\sim\frac{1}{\Lambda^{\frac{1}{2}}}
\end{equation*}
This result can be combined with our previous calculation of the amount of available entropy to obtain an estimate for the form of the probability distribution for computational capacity given $\Lambda$:
\begin{equation*}
P\left(\text{Computational Capacity}|\Lambda\right)\sim\frac{1}{\Lambda^{\frac{3}{2}}}
\end{equation*}

We should, however, think more carefully about the cost of computing and Landauer's Principle. Recalling that we produce $k_B\ \text{ln}\ 2$ entropy when we erase a bit, we only get to use each of our possible bits at most twice before we saturate the entropy bound. Furthermore, since de Sitter space is eternal\footnote{Or at least almost eternal, it seems likely that it would eventually decay quantum mechanically.} and unchanging, we have an infinite amount of time to perform calculations. Thus given we have an absolute limit on bits, the rate of ops is irrelevant, and we should use:
\begin{equation*}
P\left(\text{Computational Capacity}|\Lambda\right)\sim\frac{1}{\Lambda}
\end{equation*}
Further constraints apply when the details of causality and geometry are studied more carefully. Krauss and Starkman \cite{Krauss:2004jy} have worked this out in detail for our universe by carefully considering the process of gathering energy to use for computing. Equivalent calculations for de Sitter space change the maximum number of available bits by a numerical factor independent of $\Lambda$, and therefore do not affect $P\left(\text{Computational Capacity}|\Lambda\right)$ except for an irrelevant normalization.

While we have a robust candidate for the probability distribution of computational power as a function of $\Lambda$, it is thus far equivalent to considering solely entropy and we would like to go beyond this. Before we speculate on how we may do so, let's briefly summarize (following Bousso \cite{Bousso:2006ucsb}) the properties of the distribution we've obtained and small variations around it.

For the $1/\Lambda$ case we have (as noted above, we assume the prior distribution is uniform in the range $\Lambda_{min}$ to $\Lambda_{max}$ and that eternal inflation has no effect ($\Lambda_0$ is the measured value of $\Lambda$):
\begin{align*}
P\left(10^{-n}\Lambda_0\leq\Lambda\geq10^{n}\Lambda_0\right)&=\frac{\text{ln}\left(\frac{10^n\Lambda_0}{10^{-n}\Lambda_0}\right)}{\text{ln}\left(\frac{\Lambda_{max}}{\Lambda_{min}}\right)}\\
&=\frac{2n\ \text{ln}\,10}{\text{ln}\left(\frac{\Lambda_{max}}{\Lambda_{min}}\right)}\sim\frac{9}{\text{ln}\left(\frac{\Lambda_{max}}{\Lambda_{min}}\right)}
\end{align*}
The last line sets $n=2$ and we see that irrespective of the value of $\Lambda_0$, the probability of a $1\%$ or better tuning is about 10 times the reciprocal of logarithm of the ratio of $\Lambda_{max}$ to $\Lambda_{min}$. $\Lambda_{max}$ is presumably around 1, estimating $\Lambda_{min}$ is a little trickier, but assuming that we can ignore $\Lambda=0$ vacua -- this is not a good assumption, we'll return to it presently -- models of the landscape generally suggest $\Lambda_{min}\sim10^{-(\text{a few hundred to a thousand})}$. This gives (a very rough) estimate of the probability of being within a couple of orders of magnitude of around $1/100$ to $1/50$. Importantly this result is independent of $\Lambda_0$, so all values within the same window are equally likely (so long as $\Lambda_0>\Lambda_{min}$).

Similar calculations for different polynomial distributions give:
\begin{equation*}
P\left(\text{Computational Capacity}|\Lambda\right)\sim\frac{1}{\Lambda^{1\pm\epsilon}}\Rightarrow P\left(10^{-n}\Lambda_0\leq\Lambda\leq10^{n}\Lambda_0\right)\sim\Lambda_0^{\mp\epsilon}
\end{equation*}
We've only kept the $\Lambda_0$ dependence here, and simply note that unlike the logarithmic case the probability of measuring a cc in a given interval depends on the interval location -- in the case of computational power that grows faster with decreasing $\Lambda$ we find ourselves more likely to measure smaller values of the cc and vice versa.

\subsection*{What Else Can We Do?}
While the cc sets the maximum amount of entropy available in any given vacua, in order for those degrees of freedom to be useable for computation we need other ingredients in the universe. $\Lambda$ may provide most of the energy density, but it is not available for computation, instead we must use the degrees of freedom available in matter or radiation. In late time de Sitter space, the only matter available to observers is that which has decoupled from the Hubble flow. While there is a lot physics controlling this, the most significant factors are the cosmological constant and the size of density perturbations produced by inflation. This is, of course, very similar to Weinberg's original anthropic bound for $\Lambda$ \cite{Weinberg:1987dv} which uses the collapse fraction (the amount of matter in galaxies) to provide a probability distribution for $\Lambda$ (a later paper by Martel, Shapiro and Weinberg improves this calculation \cite{Martel:1997vi}). In late time de Sitter space we are interested in the size of each collapsed structure, rather than the collapse fraction. However, from a more refined computational point of view, both these things are important -- the latter sets the amount energy available for computing (needed if we're going to erase bits) and the former sets the number of bits we have available if we compute into the far future. In this set-up, computing early has some advantages, since there are more bits available, hence we would expect that rate of computation would also impose some constraints. Turning this into a distribution for $\Lambda$ (and for the initial size of density perturbations), is non-trivial, but certainly within the realms of possibility.

One particular problem we have ignored with regards to limits for $\Lambda$ is the presence of $\Lambda=0$ and $\Lambda<0$ vacua. In both cases this is little tricky to justify, but at least for $\Lambda<0$ one can note that such vacua cannot have long-lived cosmologies. $\Lambda=0$ (Minkowski) vacua are more of an issue, there appears to be no cosmological or computational reason to exclude them. In fact, from a computational point of view, Minkowski space is rather powerful, having an infinite amount of entropy available to an observer in the far future. One way to sidestep this rather inconvenient issue is to note that Minkowski spaces correspond to supersymmetric vacua, a fact which might constrain the computational abilities of an observer within such a space.

One further possibility for applying computational anthropic constraints is to consider the process of experiment. It is well known that physics in general, and cosmology in particular, constrain what can be measured by an observer. If such constraints can be expressed (at least partially) in terms of fundamental parameters, one might hope to further improve the anthropic measure. For example, late time de Sitter space contains very little information about early universe cosmology \cite{Krauss:2007nt}, a fact which could limit an observer's ability to compute (i.e. do physics), this in turn might imply that observers should live early in order to maximize their computational capacity, further restricting the probability distributions for ops and bits.

Rather trickier to constrain computationally are the non-gravitational parts of the theory, though there are some possibilities. To begin with the existence of matter implies our low energy theory has at least one other sector apart from gravity (though, in theory at least, matter could be provided by small black holes, which could be purely gravitational). Furthermore, even if one could build a gravitational computer, if we want to compute something specific we need to be able (at the very least) arrange a specific starting state, a process which seems like it would require an additional force (probably with a repulsive as well attractive component). With these points in mind, forces beyond gravity seem at least probable. Even with stipulated extra sectors in the theory, though, it is far from clear how their computational consequences are to be constrained, but we remain hopeful at the possibility.

\subsection*{Die Grundlagen der Physiker}
We have achieved considerably less than we would have liked\footnote{Though given the vast amount of computing power left in universe, one has hopes for future competitions.}. Starting with the lofty premise that the appropriate analogue for the Entscheidungsproblem in physics is a computational version of the anthropic principle, we found ourselves with little new to say. It turns out that while computational bounds can be placed on the spectrum of universes, these bounds are no better then just considering the entropy -- i.e. we have been unable to constrain computational complexity beyond entropy alone. However, as detailed above, there are certainly possibilities for doing better and the potential for a computational anthropic principle seems considerable.

We end where we began, with Hilbert and the foundations of physics. While the anthropic principle's implication that human mores can shape the world may be unsettling, when rephrased in the language of observation and computation it seems indisputable. And we would go further. The limits and possibilities of physics are not set by theory and mathematics alone, but rather by the fact that physics cannot just exist, it must be done. To find the foundations of physics we must look first to the foundations of physicists.

\subsubsection*{Note}
Sometime after this essay was submitted, Bousso and Harnik extended their work on the ``Entropic Principle'' \cite{Bousso:2010vi}. Though the absense of closed timelike curves limits the effect on the work contained in this essay, we would be remiss in not mentioning Bousso and Harnik's significant extension of their earlier work referenced above.

\subsubsection*{Acknowledgements}
This material is based on work support by the National Science Foundation under Grant No. PHY-0455649.

\end{document}